\documentclass[twocolumn]{aastex62ed}
\usepackage{times}
\usepackage{xspace}

\newcommand{\Msun}{\mathrm{M_{\sun}}}

\newcommand{\kms}{$\mathrm{km\,s^{-1}}$\xspace}

\def\gsim{ \lower .75ex \hbox{$\sim$} \llap{\raise .27ex \hbox{$>$}} }
\def\lsim{ \lower .75ex \hbox{$\sim$} \llap{\raise .27ex \hbox{$<$}} }

\newcommand{\beginsupplement}{%
        \setcounter{table}{0}
        \renewcommand{\thetable}{S\arabic{table}}%
       \renewcommand{\thefigure}{S\arabic{figure}}%
         \setcounter{section}{0}
        \renewcommand{\thesection}{S\arabic{section}}%

     }

\graphicspath{{./}{figures/}}

\received{2019 March 14}
\revised{2019 March 26}
\accepted{2019 March 26}
\submitjournal{ApJL}

\shorttitle{Dark matter decay in Perseus}
\shortauthors{Lovell et al.}

\begin{document}

\title{\bf Simulating the dark matter decay signal from the Perseus galaxy cluster}

\correspondingauthor{Mark R. Lovell}
\email{lovell@hi.is}

\author[0000-0001-5609-514X]{Mark R. Lovell}
\affiliation{University of Iceland, Dunhagi 5, 107 Reykjav\'ik, Iceland}
\affiliation{Durham University, South Road, Durham, DH1 3LE, UK}
\author[0000-0002-6969-0738]{Dmytro Iakubovskyi}
\affiliation{Bogolyubov Institute of Theoretical Physics, Metrologichna Str. 14-b, 03143, Kyiv, Ukraine}
\author[0000-0002-6583-9478]{David Barnes}
\affiliation{Department of Physics, Kavli Institute for Astrophysics and Space Research, Massachusetts Institute of Technology, Cambridge, MA 02139, USA}
\author[0000-0002-0974-5266]{Sownak Bose}
\affiliation{Harvard-Smithsonian Center for Astrophysics, 60 Garden Street, Cambridge, MA 02138, USA}
\author[0000-0002-2338-716X]{Carlos S. Frenk}
\affiliation{Durham University, South Road, Durham, DH1 3LE, UK}
\author[0000-0002-3790-9520]{Tom Theuns}
\affiliation{Durham University, South Road, Durham, DH1 3LE, UK}
\author[0000-0003-4634-4442]{Wojciech A. Hellwing}
\affiliation{Center for Theoretical Physics, Polish Academy of Sciences, Aleja Lotnik\'ow 32/46, 02-668 Warsaw, Poland}

\begin{abstract}

The nearby Perseus galaxy cluster is a key target for indirect
detection searches for decaying dark matter. We use the C-EAGLE
simulations of galaxy clusters to predict the flux, width and shape of
a dark matter decay line, paying particular attention to the
unexplained 3.55~keV line detected in the spectra of some galaxies and
clusters, and the upcoming {\it XRISM} X-ray observatory mission. We show
that the line width in C-EAGLE clusters similar to Perseus is
typically [600-800]~\kms, and therefore narrower than the amplitude
of the velocity dispersion of galaxies in the cluster. Haloes that are
significantly disturbed can, however, exhibit galaxy velocity
dispersions higher than 1000~\kms, and in this case will show a large
difference between the line profiles of on- and off-center
observations. We show that the line profile is likely to be slightly
asymmetric, but still well approximated by a Gaussian at the 10\%
level, and that the halo asymmetry can lead to fluxes that vary by a
factor of two. In summary, we predict that, if the previously reported
3.55~keV line detections do originate from dark matter decay, the
{\it XRISM} mission will detect a line with a roughly Gaussian profile at a
rest frame energy of 3.55~keV, with a width $>600$~\kms and
flux approximately in the range $[4-9]\times10^{-8}\mathrm{counts/sec/cm^{2}}$. 

\end{abstract}


\section{Introduction} \label{sec:intro}

One of the possible means for identifying dark matter is the detection of photons that are emitted during the annihilation or decay of dark matter particles in astrophysical objects such as galaxies and clusters of galaxies. The decay channel is favoured if the dark matter particle is light ($m_\mathrm{p}c^{2}\ll$~GeV). An unidentified line feature at an energy of 3.55~keV has been reported in multiple astronomical objects \citep[e.g.][]{Bulbul14,Boyarsky14a,Cappelluti17,Boyarsky19}. One of the possible origins for this line is the decay of a 7.1~keV-mass dark matter particle.

In order to confirm or exclude dark matter decay as the origin for these 3.55~keV photons, multiple targets will have to be observed with X-ray observatories, and their flux amplitudes measured  and found to all be consistent with originating from a particle of the same mass and lifetime, as was shown in \cite{Lovell19a}. In that paper, one of the most promising targets was the Perseus galaxy cluster, which has the benefit that its predicted line has not two but three properties that can be measured: its flux, its energy and its line width, the last of which is broad enough to have been resolved by the defunct {\it Hitomi} mission \citep{HitomiC16,Tamura19}, and will be resolved by the upcoming {\it XRISM} ($\sim$2022; $\sim5$~eV energy resolution) and {\it ATHENA/XIFU} ($\sim$2028; $\sim2.5$~eV energy resolution) missions, and the proposed {\it Lynx/LXM} observatory ($>$2030;$\sim3$~eV energy resolution). Having a model that describes the height, width and shape of the Perseus cluster is a crucial component of the search for decaying dark matter. This is complicated, however, by the fact that the  dark matter velocity dispersion, and thus the line width, cannot be measured directly from observations, and the most popular proxy for this quantity, the velocity dispersion of the cluster member galaxies (as considered in \citealp{HitomiC16}), may not be sufficiently accurate for our purposes \citep[e.g.][]{Armitage18,Elahi18}.

In this {\it Letter} we make predictions for the flux, equivalent width, and shape of the hypothesised dark matter decay line in Perseus using the method based on hydrodynamical cosmological simulations of \cite{Lovell15,Lovell19a}, with particular attention to the differences between the member galaxy and dark matter velocity dispersions, and also to the degree of deviation of the line shape from a Gaussian. In Section~\ref{sec:methods} we review our simulation suite and the method from \cite{Lovell19a}, then present our results in Section~\ref{sec:results} and draw conclusions in Section~\ref{sec:conclusions}.  

\section{Simulations and methods} \label{sec:methods}

We use the thirty hydrodynamical simulations of cluster zooms included in the C-EAGLE project, which are described in full in \cite{Bahe17} and \cite{Barnes17}. Briefly, the galaxy formation model is the AGNdT9 version of the EAGLE model \citep{Schaye15}, which features cooling, star formation, supernova feedback and black hole growth and feedback; the AGNdT9 model is optimized over the standard EAGLE reference model to improve the gas--to--total mass fractions and X-ray luminosity--temperature relations in massive clusters. The dark matter particle mass is $9.7\times10^{6}\Msun$, the softening length is 0.7~kpc at $z=0$, and the cosmological parameters were chosen to be consistent with the \cite{PlanckCP13} results: Hubble parameter, $h = H_{0}/(100~\mathrm{km~s^{-1}})$ = 0.6777; dark energy density, $\Omega_{\Lambda}=0.693$; matter density $\Omega_\mathrm{M} = 0.307$; and baryon energy density $\Omega_\mathrm{b}= 0.04825$. 

The method for determining the dark matter decay flux is the same as that first presented in \cite{Lovell15} and expanded in \cite{Lovell19a}. To summarize, we place an observer at a distance of 69.5~Mpc -- i.e. the distance to Perseus -- from the center of the most massive galaxy in each simulated cluster, draw a cone determined by the field-of-view (FoV) of the {\it XRISM} telescope (which we treat as a circle of radius $1.4'$) and the observer--cluster axis (radius 28~kpc at the cluster center); we have also performed these analyses with the {\it ATHENA/XIFU} FoV ($2.5'$; 50~kpc aperture radius at the Perseus distance), which is very similar in its FoV and spectral resolution to the proposed {\it Lynx/LXM} instrument.  Having chosen an FoV, we identify the dark matter particles located within the cone delineated by that FoV. We treat each dark matter particle as a point source of decay photons emitted isotropically at a constant rate. 

The decay flux is then the total flux measured from dark matter particles within the FoV; we do not add any contribution from intervening dark matter unassociated with the cluster. The shape of the line -- and thus its dispersion and full width-half-maximum (FWHM) -- are calculated by binning the flux from each particle in line-of-sight (l.o.s., i.e. 1D) velocity. We repeat this process for 500 observer locations distributed at random on a sphere of radius 69.5~Mpc. For each halo we select our galaxy and dark matter particle locations from the $z=0$ simulation snapshot, which is a good approximation for the Perseus cluster ($z=0.0167$). The one difference from the mock Perseus observations reported in \cite{Lovell19a} is that we select particles up to 10~Mpc from the cluster center for analysis, as opposed to 2~Mpc in that previous study, to ensure we include the contribution to the line profile of any high velocity infalling dark matter.

We calculate the 1D cluster galaxy velocity dispersion in a way that mirrors the procedure of \cite{Tamura14}, who used CfA and CMASS redshifts for 100 galaxies within $30'$ of the center of Perseus. Using the same 500 observer positions calculated for the X-ray flux mock observations, we draw a cone defined by the cluster center/observer axis and an opening angle of $30'$ (radius 600~kpc at the halo center, which is a third of the virial radius). We select the first 100 galaxies in the halo/subhalo catalog, ranked by halo mass, whose center-of-potential is located within this cone and compute their 1D velocity dispersion along the line-of-sight associated with that observer. For each of the 500 observers we therefore measure two velocity dispersions ($\sigma_\mathrm{1D}$): one for the dark matter particles, $\sigma_\mathrm{1D,DM}$, and one for the member galaxies, $\sigma_\mathrm{1D,gals}$. 

We repeat the $\sigma_\mathrm{1D,gals}$ procedure for the observed Perseus galaxies of the \cite{Tamura14} sample, and obtain an observed velocity dispersion of 1210~\kms if the bright, CMASS galaxies are used and 1330~\kms if the CfA galaxy sample is used, both of which are in broad agreement with the older measurement of \cite{Kent83}. For the purposes of this paper we take the approximate limit on the value of the observed 1D velocity dispersion of Perseus galaxies to be [1200,1400]~\kms. We will compare these values specifically to those haloes that have the same measured virial mass, $M_{200}$, as Perseus, where $M_{200}$ is defined as the mass within the radius that encloses an overdensity 200 times the critical density of the Universe. We adopt the Perseus $M_{200}$ to be $6.65^{+0.43}_{-0.46}\times10^{14}\Msun$ \citep{Simionescu11}, and the haloes that match this $M_{200}$ range are haloes 18 ($6.94\times10^{14}\Msun$) and 19 ($6.84\times10^{14}\Msun$). Finally, we repeat parts of this analysis for three further clusters -- Ophiuchus, Virgo and Centaurus -- and present those results in the supplementary material.     

\section{Results} \label{sec:results}

\begin{figure}
   \includegraphics[scale=0.34]{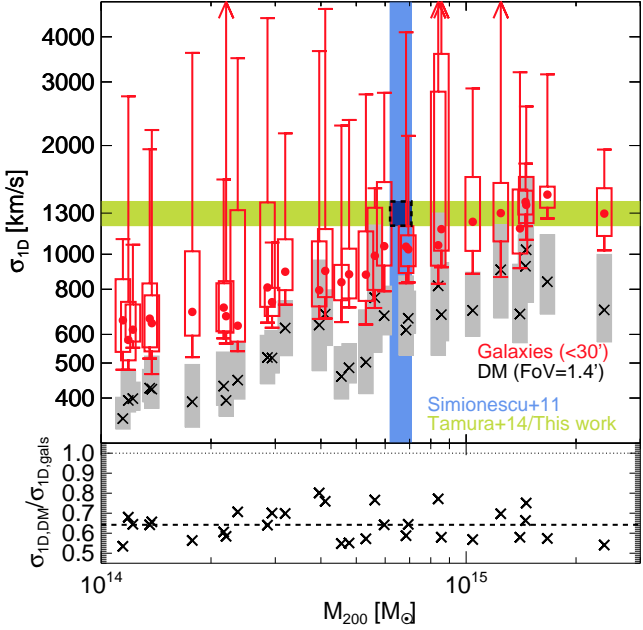}
    \caption{{\it Top panel:} C-EAGLE cluster dark matter and galaxy velocity dispersions as a function of halo mass. The median values of $\sigma_\mathrm{1D,DM}$ over 500 sightlines are plotted as black crosses, and the region encompassing 95\% of the data for each halo as a gray rectangle. Similarly, the median $\sigma_\mathrm{1D,gals}$ are shown as red circles. The red boxes show the 68\% region in  $\sigma_\mathrm{1D,gals}$ for each halo, and the error bars the 95\% region; in cases where the upper 95\% limit is above 5000~\kms, that upper limit is truncated and represented by an arrow. The Perseus value of $\sigma_\mathrm{1D,gals}$ that we measure using the \cite{Tamura14} data is plotted as a green band. The blue band is the measured range of $M_{200}$ for Perseus $(6.19-7.02)\times10^{14}\Msun$. {\it Bottom panel:} the ratios of the median $\sigma_\mathrm{1D,DM}$ and $\sigma_\mathrm{1D,gals}$. The median ratio of $\sigma_\mathrm{1D,DM}$ relative to $\sigma_\mathrm{1D,gals}$ is shown as a dashed line.}
    \label{fig:v1d}
\end{figure}

We begin the presentation of our results by showing the differences between $\sigma_\mathrm{1D,DM}$ and $\sigma_\mathrm{1D,gals}$ for clusters across the mass range $1\times10^{14}-2\times10^{15}\Msun$, and then how those with the Perseus mass compare to the observationally measured Perseus $\sigma_\mathrm{1D,gals}$. In Fig.~\ref{fig:v1d} we plot $\sigma_\mathrm{1D,DM}$ and $\sigma_\mathrm{1D,gals}$ for all 30 C-EAGLE haloes as a function of $M_{200}$. 

The difference between the values of the two sets of velocity dispersion is significant. At all halo masses, the median $\sigma_\mathrm{1D,gals}$ is about 70\% higher than $\sigma_\mathrm{1D,DM}$\footnote{The scatter is also much higher for $\sigma_\mathrm{1D,gals}$ than for $\sigma_\mathrm{1D,DM}$: this is due to the sampling error difference between taking 100 galaxies vs. millions of dark matter particles.}, Clearly, taking the \cite{Kent83} $ \sigma_\mathrm{1D,gals}=1300$~\kms -- the value for $\sigma_\mathrm{1D,gals}$ adopted by \cite{HitomiC16} -- as a proxy for the velocity dispersion of a dark matter decay line is a large overestimate. Our calculation also shows that the $\sigma_\mathrm{1D,gals}$ in the C-EAGLE simulations are in good agreement with the value inferred from the CMASS and CfA data. This high value is similar to that found for the same data sets by \cite{Armitage18b}, who also showed that cluster galaxy orbits are preferentially radial rather than tangential; a similar result for a different data set was shown by \cite{Elahi18}.

We now check whether there is any correlation between $\sigma_\mathrm{1D,gals}$ and $\sigma_\mathrm{1D,DM}$ for each halo, by calculating the Spearman's rank correlation coefficient across the 500 sightlines for each of the 30 haloes. We then test the significance of these coefficients by comparing the results to 10,000 randomly drawn sets of 500 ranked-pairs, and also by applying a Fisher transformation/z-score test. Under both tests, only four of the thirty haloes showed Spearman coefficients consistent with the null hypothesis that there is no correlation between $\sigma_\mathrm{1D,gals}$ and $\sigma_\mathrm{1D,DM}$ at three standard deviations. Twenty-five showed a significant preference for a positive correlation and the last one showed a strong preference for an {\it anti}-correlation. 

We focus the rest of our analysis on those sightlines that best describe Perseus as per our constraints. These are the sightlines that run through Perseus-mass haloes and have the measured Perseus galaxy velocity dispersion, i.e. those that are  enclosed by dashed lines at the intersection of the bands in Fig.~\ref{fig:v1d}. We thus choose those sightlines that belong to the two haloes that are within the measured mass range for Perseus -- haloes 18 and 19 -- and have $\sigma_\mathrm{1D,gals}$ in the range [1200,1400]~\kms, of which there are only 37 since the measured Perseus $\sigma_\mathrm{1D,gals}$ is significantly above the median simulated $\sigma_\mathrm{1D,gals}$ at this halo mass range. We dub these `Perseus-analog' sightlines; however, the value of the Spearman correlation coefficient in $\sigma_\mathrm{1D,gals}-\sigma_\mathrm{1D,DM}$ for these two haloes is only 0.37 and 0.26, so our results are not strongly influenced by this choice.

\begin{figure}
   \includegraphics[scale=0.34]{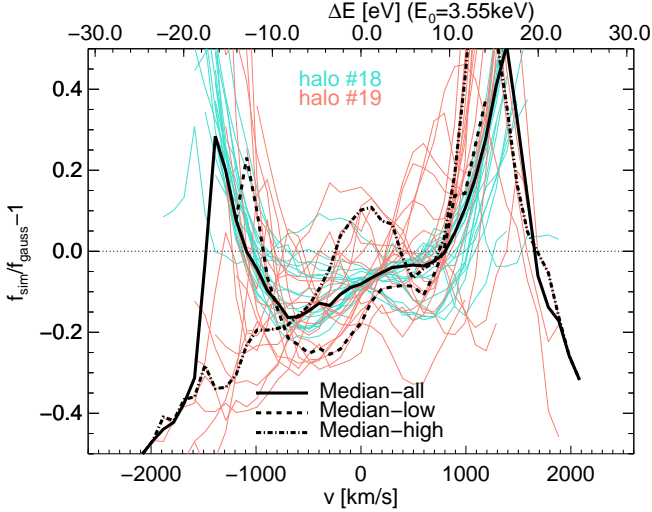}
    \caption{Ratio of the difference between each Perseus-analog velocity profiles relative to a Gaussian line with the same FWHM and total flux amplitude. Individual sightlines are plotted as thin lines and are colored according to host halo (turquoise and pink). Median relations are plotted as thick black lines: a solid line for the whole sample, dashed for the high-ratio subsample and dot-dashed for the low ratio subsample. We plot the velocities  on the lower $x$-axis and energies on the upper $x$-axis (both rest frame), where the energy scale assumes an emission energy $E_{0}=3.55$~keV. The velocity/energy resolution of the {\it XRISM} instrument is shown in the top left corner.}
    \label{fig:SvG}
\end{figure}

The total flux of the line is anticipated from the mass and, to a lesser extent, the concentration of the Perseus halo (as computed, for example, in \citealp{Lovell19a}), while the line width can be estimated from the virial theorem alongside assumptions about the anisotropy of dark matter particles orbits. The line shape is typically assumed to be a Gaussian. We test the degree of deviation from the Gaussian in the following manner. For each of our subsamples of Perseus-analog sightlines we compute the unique Gaussian that has the same total flux and FWHM as each sightline and then calculate the ratio of the difference between the simulation sightline and the Gaussian approximation, i.e.:

\begin{equation}
    \mathrm{ratio}(v) = \frac{f_\mathrm{sim}(v)-f_\mathrm{gauss}(v)}{f_\mathrm{gauss}(v)}
\end{equation}

\noindent
where $f_\mathrm{sim}(v)$ and $f_\mathrm{gauss}(v)$ are the line profiles measured from the simulation and using the Gaussian approximation respectively. We also reflect each Perseus-analog curve about velocity $v=0$ if necessary to ensure that there is more flux at positive velocities than at negative, thus maximizing the visibility of any line asymmetry present. In Fig.~\ref{fig:SvG}, we plot each line difference ratio as a thin line. We also plot three sets of median relations defined by the value of the ratio at the velocity, $v=0$~\kms: the $\sim15$\% that have the highest ratio value at $v=0$~\kms and also the $\sim15$\% lowest. 

Within 1000~\kms either side of the line centroid, the deviation from the Gaussian line profile is typically less than 20\%. The median differences of the subsample are deficits of 10\% at the centroid, 5\% at $v=+700$~\kms and 20\% at $v=-700$~\kms, suggesting that any asymmetry of the line could be detectable with the {\it XRISM} energy resolution. These values are suppressed by a further 5-10\% for the low-peak subsample, and correspond to simulated curves that have broader tails than the Gaussian as show by the sharply increasing ratio values for velocity $|v|>1000$~\kms. This excess of flux at large velocities is also apparent in the high-peak subsample, where the peak value is $10$\% higher than the Gaussian but otherwise follows the pattern of a deficit at $|v|=700$~\kms and an enhancement at larger velocities. We note that there is an apparent difference between the two haloes, with the Gaussian providing a better match to the halo 18 curves than to those of halo 19; this is likely because halo 19 is less relaxed than halo 18 \citep{Barnes17}. Finally, we have repeated this exercise with a series of sightlines offset from the halo center by $1'$, as used for some of the {\it Hitomi} observations, and find no significant differences to our on-center results. We conclude that a Gaussian is an appropriate approximation to the proposed Perseus dark matter decay line at the 20\% level, and otherwise the biggest difference is in the presence of broader tails in the distribution. For the remainder of this letter, we  use the $\sigma_\mathrm{1D,DM}$ measured directly from the simulations, and not that from a Gaussian fit.     

\begin{figure}
  \includegraphics[scale=0.55]{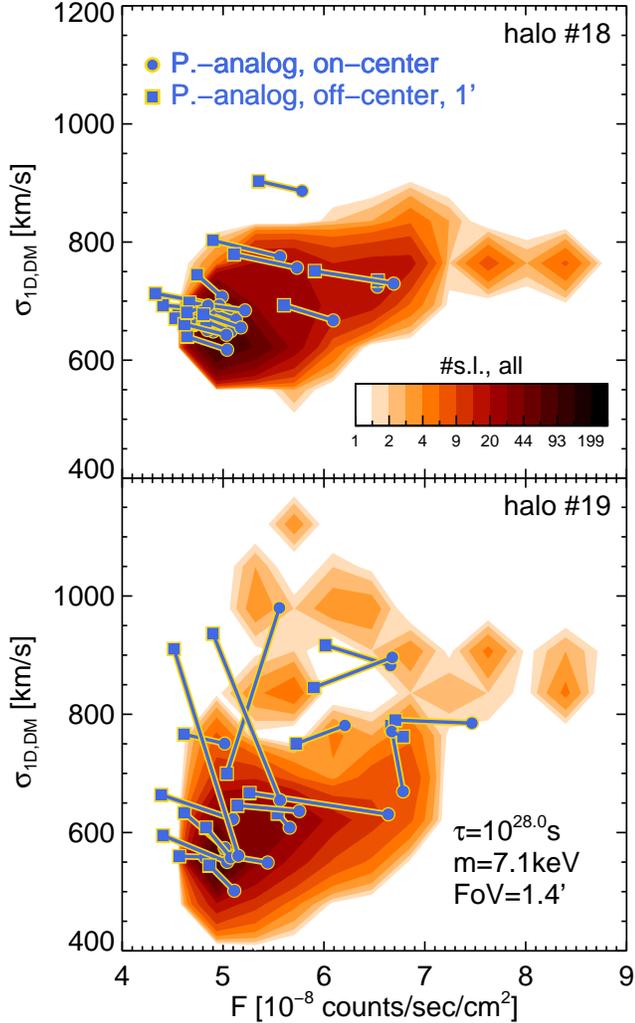}
    \caption{Heat map plots in $\sigma_\mathrm{1D,DM}$ and in flux, $F$ for all 500 halo~18 (top panel) and halo~19 (bottom panel) on-center sightlines (s.l.). The color bar indicates the number of s.l. in each contour cell, and we use the same color bar for both panels. The Perseus-analog sightlines are shown as blue circles. $1'$ off-center counterparts to the Perseus-analog sightlines are shown as blue squares, and are linked to their corresponding on-center observations with solid lines.}
    \label{fig:XRISMcontour}
\end{figure}

 We now present some predictions for the properties of the line to be probed by the {\it XRISM} satellite: $\sigma_\mathrm{1D,DM}$ and flux, $F$. We present our results as a series of contours in $\sigma_\mathrm{1D,DM}$ and $F$ for mock observations made with the {\it XRISM} FoV (using a circular aperture of radius $1.4'$), assuming a dark matter particle mass of 7.1~keV and a particle lifetime of $1\times10^{28}$~sec \footnote{Note that the flux is inversely proportional to both the particle mass and the lifetime but the velocity dispersion is independent of these parameters}. We draw one set of contours for each of the haloes that we consider, using all 500 on-center sightlines, and plot the 37 Perseus-analog on-center sightlines as circles. We also include the Perseus-analog sightlines from our off-center ($1'$) observations, which is the off-center angle in some of the {\it Hitomi} observations, as squares. Each off-center mock observation was performed from the same observer location as one of the on-center observations, and we indicate which off-center--on-center pairs share a common observer position using solid lines. We present these results in Fig.~\ref{fig:XRISMcontour}.

Both haloes' $F$--$\sigma_\mathrm{1D,DM}$ distributions are highly asymmetric, with a preference for $F\sim5\times10^{-8}\mathrm{counts/sec/cm^{2}}$ and a long tail to fluxes almost twice as large. There is also a slight preference for a correlation between $F$ and $\sigma_\mathrm{1D,DM}$. The spread in $\sigma_\mathrm{1D,DM}$  values is twice as large for the less relaxed halo~19 as for the more relaxed halo~18; for both haloes $\sigma_\mathrm{1D,DM}$ is significantly more than the 180~\kms dispersion derived for the hot cluster gas in \cite{HitomiC16}. Differences in halo formation history / asymmetry are important for the properties of the signal. Evidence for this difference between relaxed and unrelaxed haloes comes in the relationship between the on- and off-center observations of Perseus-analog sightlines. The halo~18 subsample shows a remarkably regular preference for the off-center sightlines to have lower flux and slightly higher $\sigma_\mathrm{1D,DM}$ than their on-center counterparts, with a median $3.9\times10^{-9} \mathrm{counts/sec/cm^{2}}$ decrement in flux and 18~\kms enhancement in $\sigma_\mathrm{1D,DM}$. By contrast, the halo~19 Perseus-analog sightlines show a much greater range of behaviours, from increases in $\sigma_\mathrm{1D,DM}$ of 350~\kms from on-center to off-center, to the same change in the opposite direction. An in-depth prediction would require constrained simulations of Perseus-analog haloes that are beyond the scope of this letter. Finally, we have repeated this exercise using the {\it ATHENA/XIFU} FoV, which we approximate as a circle of radius $2.5'$. We obtain results similar to those for {\it XRISM}, with the following exceptions: the range of flux is $[1.1-2.1]\times10^{-7}\mathrm{counts/sec/cm^{2}}$; the high-flux tail for halo~18 shows an increased  $\sigma_\mathrm{1D,DM}\sim870$~\kms and is therefore the same as that of halo~19; and the differences between the on- and off-center observations shrink in relative terms, both for $F$ and for $\sigma_\mathrm{1D,DM}$. 

\section{Conclusions} \label{sec:conclusions}

If the dark matter is a particle that decays, the Perseus galaxy cluster is an excellent target for which to collect data suitable for indirect detection. The interpretation of the observations requires an estimate of the decay line amplitude, width and shape. In this letter, using the C-EAGLE cosmological hydrodynamical simulations of massive clusters, we have calculated these three line properties and explore how they correlate with the Perseus halo properties. 

We showed that the 1-D velocity dispersion of the dark matter -- as measured within the {\it Hitomi/XRISM} field-of-view (FoV) -- is typically 40\% smaller than the 1-D velocity dispersion of the cluster member galaxies when evaluated using members within $30'$ of the cluster center (Fig.~\ref{fig:v1d}). We showed that there is a correlation between the two velocity dispersions, but it is very weak and not present for all 30 simulated haloes. We confirmed that the line profile is well described by a Gaussian within 1000~\kms of the line center, although the presence of line asymmetry is common (Fig.~\ref{fig:SvG}). We also presented estimates of the distributions of fluxes and dark matter velocity dispersions for different lines-of-sight, and found that both could vary by as much as a factor of two. This is particularly true if the halo is unrelaxed, in which case there can be dramatic differences between the line profiles on- and off-center (Fig.~\ref{fig:XRISMcontour}). These differences are much smaller when the larger {\it ATHENA/XIFU} FoV is used; otherwise the results from that instrument are similar to those for the {\it XRISM} FoV.

In summary: we have shown that the velocity dispersion of the proposed dark matter decay line in Perseus is much smaller than the velocity dispersion of cluster galaxies. We predict that, if the reported 3.55~keV line does indeed originate from dark matter decay, the upcoming observations of Perseus by the {\it XRISM} mission -- assuming an exposure time of 1~Msec \citep{Bulbul14} -- will detect a line with the following properties: a total flux of $[4,9]\times10^{-8}\mathrm{counts/sec/cm^{2}}$, a velocity dispersion of $[600,800]$~\kms -- although exceptionally as large as 1200~\kms -- and a shape that is within 20\% of Gaussian.

\section*{Acknowledgments}

 MRL is supported by a COFUND/Durham Junior Research Fellowship under EU grant 609412 and also by a Grant of Excellence from the Icelandic Research Fund (grant number 173929−051). DI acknowledges support from the grant for young scientist’s research laboratories of the National Academy of Sciences of Ukraine. CSF acknowledges a European Research Council Advanced Investigator grant DMIDAS (GA 786910) and support for the STFC Consolidated Grant for Astronomy at Durham. This work used the DiRAC@Durham facility managed by the Institute for
Computational Cosmology on behalf of the STFC DiRAC HPC Facility
(www.dirac.ac.uk). The equipment was funded by BEIS capital funding
via STFC capital grants ST/K00042X/1, ST/P002293/1, ST/R002371/1 and
ST/S002502/1, Durham University and STFC operations grant
ST/R000832/1. DiRAC is part of the National e-Infrastructure. WAH is supported by an Individual Fellowship of the Marie Sk\l odowska-Curie Actions and therefore acknowledges that this project has received funding from the European Union's Horizon 2020 research and innovation program under the Marie Sk\l odowska-Curie grant agreement No 748525.

\bibliographystyle{yahapj}

\newpage

\section*{Supplementary Material}
\beginsupplement

In this supplementary information we replicate Figs.~\ref{fig:v1d} and \ref{fig:XRISMcontour} of our Perseus analysis for three other clusters: Virgo, Centaurus and Ophiuchus. We choose these three clusters because they were used as part of some prominent subsamples of the \cite{Bulbul14} analysis. All of the methodology applied is the same as Perseus, including a $30'$ opening angle / 100 member galaxy sample for the cluster galaxy velocity dispersion. Another cluster of interest is Coma; however, the measured mass and redshift for Coma are very similar to that of Perseus, so the results are likely to be very similar to our Perseus results and therefore we do not repeat the analysis for Coma. 

\section{Ophiuchus}

\begin{figure}[h]
    \centering
   \includegraphics[scale=0.34]{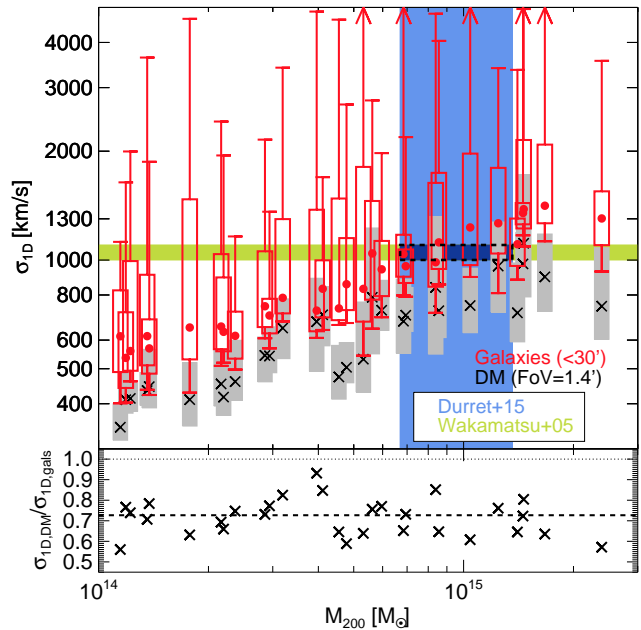}
    \caption{Repetition of Fig.~\ref{fig:v1d} for haloes placed at the Ophiuchus redshift, where the observations are given by the measured Ophiuchus values.}
    \label{fig:v1dOph}
\end{figure}

Ophiuchus is one the closest large ($M>10^{15}\Msun$) galaxy clusters to the Milky Way. For our purposes, Ophiuchus is a cluster with a redshift $z=0.028$ ($\sim120$~Mpc, or 75~per~cent further away than Perseus), a mass $M_{200}=11^{+4.3}_{-2.6}\times10^{14}\Msun$ \citep{Durret15}, and a velocity dispersion of $\sigma_\mathrm{1D,gals}=[1000,1100]$~\kms; the velocity dispersion was calculated using galaxies within 30' of the cluster center, just like Perseus \citep{Wakamatsu05}. We present the results for our clusters located at $z=0.028$ in Fig.~\ref{fig:v1dOph}; we adopt the {\it XRISM} FoV. 

\begin{figure}[h]
  \centering
  \includegraphics[scale=0.55]{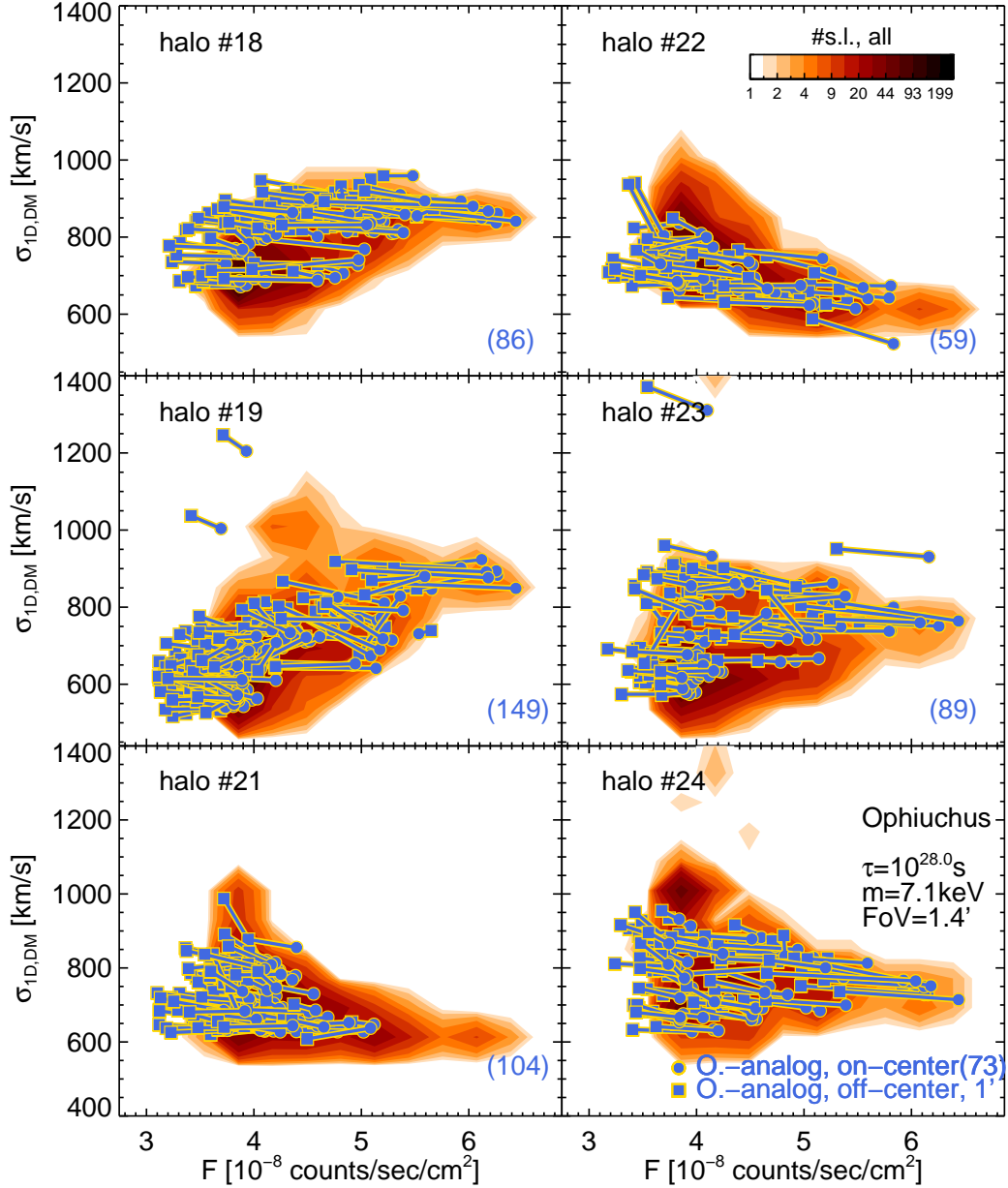}
    \caption{Repetition of Fig.~\ref{fig:XRISMcontour} for Ophiuchus, using the six haloes that match the Ophiuchus $M_{200}$; each halo is presented in a separate panel. The number of $\sigma_\mathrm{1D,gals}$-selected sightlines in each panel is shown in brackets.}
    \label{fig:XRISMcontourPhi}
\end{figure}

Unlike Perseus, the measured Ophiuchus galaxy velocity dispersion is well within the 68~per~cent region for the simulated $\sigma_\mathrm{1D,gals}$ in the allowed mass range. The difference between the galaxy and dark matter velocity dispersions is smaller than was the case for Perseus, with an average ratio of 30\%: we expect this is because the cluster is 75\% further away than Perseus and so the 30' FoV encompasses more galaxies with low (l.o.s.) relative velocities in the plane of the sky. There are six haloes that fall within the measured mass range for Ophiuchus; we plot the $\sigma_\mathrm{1D,DM}$--$F$ contours for these six haloes in Fig.~\ref{fig:XRISMcontourPhi}.  

The six haloes consistently show a preference for $F\sim4\times10^{-8}\mathrm{counts/sec/cm^{2}}$, with a long tail that extends to $6\times10^{-8}\mathrm{counts/sec/cm^{2}}$. The dark matter velocity dispersion is much more variable: although all six haloes present $\sigma_\mathrm{1D,DM}$ in the range [600,1000]~\kms, some haloes show a positive correlation between the two quantities (haloes 18 and 19), whereas 22 presents a negative correlation. Haloes 23 and 24 also present a small subset of $\sigma_\mathrm{1D,DM}>1200$~\kms. In general the $\sigma_\mathrm{1D,gals}$-selected sightlines show similar behaviour to Perseus, with off-center fluxes suppressed at the 10s of per~cent level compared to their on-center counterparts, with the off-center $\sigma_\mathrm{1D,DM}$ values typically enhanced at the $<10$\% level. With the possible exception of low flux-high dispersion sightlines, the $\sigma_\mathrm{1D,gals}$ subsample is not consistently biased relative to the full 500 sightlines.   

\section{Virgo}

\begin{figure}[h]
    \centering
   \includegraphics[scale=0.34]{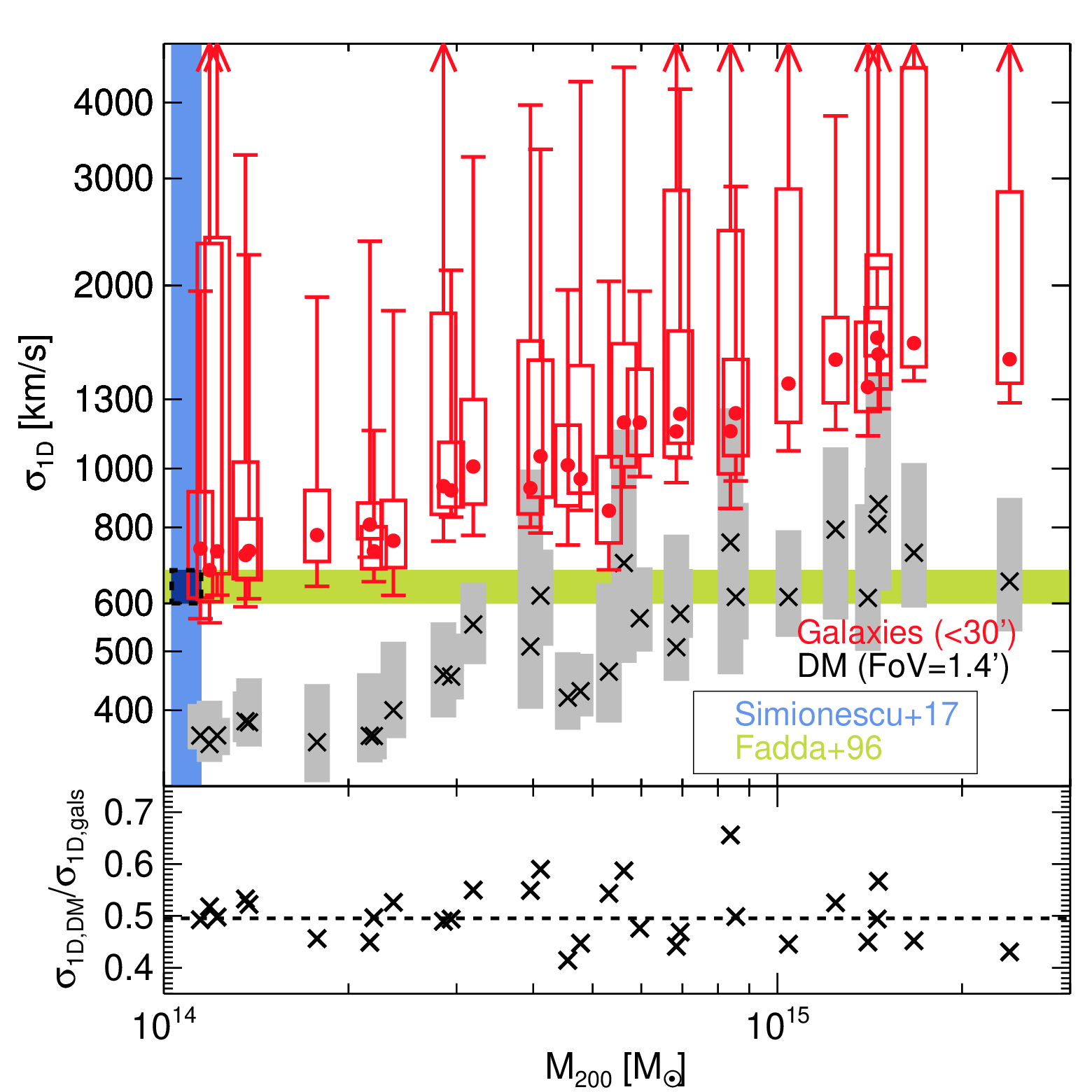}
    \caption{Repetition of Fig.~\ref{fig:v1d} analysis for the clusters at the Virgo redshift. The Virgo value of $\sigma_\mathrm{1D,gals}$ measured by \cite{Fadda96} is shown as a green band, and the blue band is the allowed range of $M_{200}$ for Virgo: $M_{200}=(1.03-1.07)\times10^{14}\Msun$ \citep{Simionescu17}.}
    \label{fig:v1dVirgo}
\end{figure}

The Virgo cluster is the closest cluster to the MW, and thus an important candidate for dark matter indirect detection. It is located at a redshift $0.0036$ ($\sim16$~Mpc, 22\% of the distance to Perseus), the galaxy velocity dispersion is $632^{+41}_{-29}$~\kms \citep[$\sim500$ member galaxies; ][]{Fadda96} and the halo mass $M_{200}=1.05\pm0.02\times10^{14}\Msun$ \citep{Simionescu17}. Our comparison between C-EAGLE cluster velocity dispersion at the Virgo distance and the Virgo data is shown in Fig.~\ref{fig:v1dVirgo}. 

The difference between the galaxy and dark matter velocity dispersions is stronger than was the case for Perseus, with the galaxy velocity dispersion on average a factor of two larger than the dark matter velocity dispersion. Virgo is therefore in some ways the opposite of Ophiuchus: it is much closer than Perseus, so our 30' subtends a smaller part of the halo radius and thus only selects the high l.o.s velocity galaxies, given that the galaxy orbits are preferentially radial. The measured Virgo velocity dispersion is located squarely within the 68~per~cent region of our three lowest mass clusters. None of the clusters has a mass within the $1\sigma$ error bar derived by \cite{Simionescu17}; we proceed with taking just the lowest mass halo in the sample, halo~1, to be a Virgo analog; for a discussion of halo-to-halo scatter we refer the reader back to the section on Ophiuchus.   

\begin{figure}[h]
    \centering
  \includegraphics[scale=0.34]{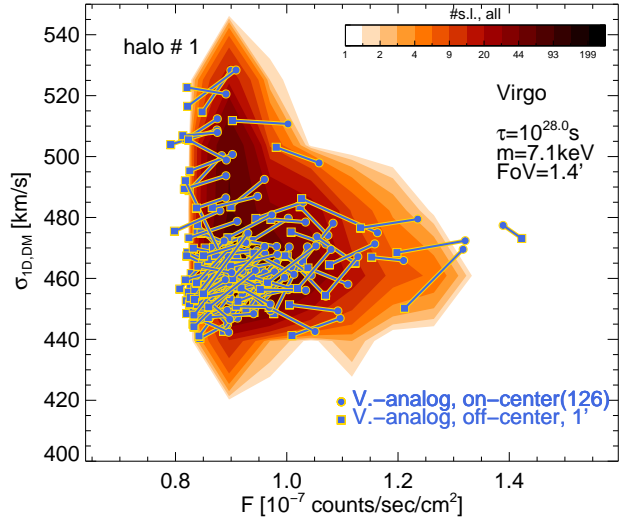}
    \caption{Repetition of Fig.~\ref{fig:XRISMcontour} for Virgo, using halo~1 from our sample placed at the Virgo redshift.}
    \label{fig:XRISMcontourVirgo}
\end{figure}

The most likely observed dark matter decay flux for Virgo, as modeled by halo~1, is $\sim0.9\times10^{-7}\mathrm{counts/sec/cm^{2}}$, and the high-flux tail extends to $\sim1.4\times10^{-7}\mathrm{counts/sec/cm^{2}}$. The velocity dispersion sits in the range [420,540]~\kms. By comparing to the Ophiuchus results in Fig.~\ref{fig:XRISMcontourPhi}, we expect that similar haloes of the same mass may present similar similar ranges in flux and velocity dispersion but potentially with a population of high flux -- high velocity dispersion sightlines that are not present for halo~1. The scatter between the velocity dispersions of the $\sigma_\mathrm{1D,gals}$ selected sample, relative to the change in flux, is bigger than for Ophiuchus, which likely reflects the extra variation between sightlines when using an aperture that is small relative to the angular size of the cluster.

\section{Centaurus}

\begin{figure}[h]
    \centering
   \includegraphics[scale=0.34]{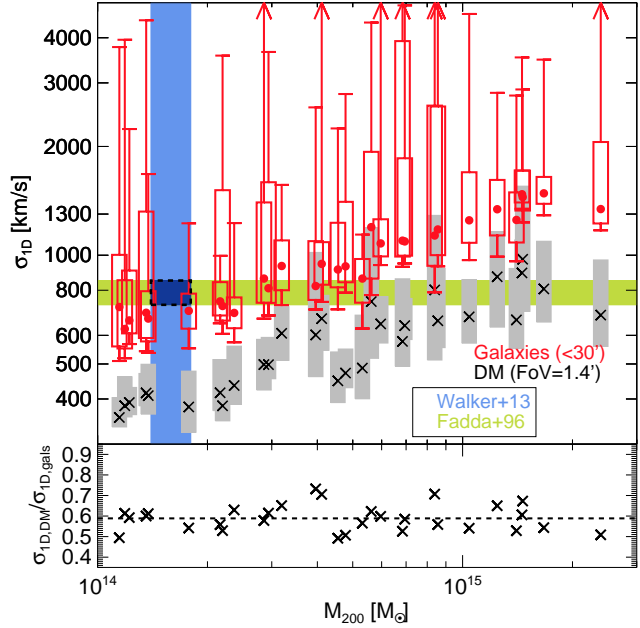}
    \caption{Repetition of the Fig.~\ref{fig:v1d} analysis for the C-EAGLE clusters when placed at the redshift of Centaurus ($z=0.0109$).}
    \label{fig:v1dCentaurus}
\end{figure}

Finally, we present results for the Centaurus cluster, which is located at a redshift $z=0.0109$ ($\sim48$~Mpc, 68\% of the distance to Perseus). It has a measured mass $M_{200}=1.6^{+0.3} _{-0.2}\times10^{14}\Msun$ \citep{Walker13}. The velocity dispersion measurement is complicated by the apparent presence of two peaks in the velocity distribution \citep{Fadda96}; we adopt the velocity dispersion that \cite{Fadda96} derived for this cluster $791^{+60}_{-62}$~\kms and comment below on the possible impact of substructure. The comparison between the galaxy and dark matter velocity dispersions is presented in Fig.~\ref{fig:v1dCentaurus}.

The dark matter velocity dispersions are suppressed by 40\% relative to the galaxy velocity dispersions, which is similar to Perseus. The measured $\sigma_\mathrm{1D,gals}$ is in agreement with the simulation predictions, although at the higher end of the allowed region which is consistent with the presence of a small subcluster along the line of sight. Only one of our haloes -- halo~4 -- has a mass within the measured range for Centaurus: we adopt this one halo as our  Centaurus analog, and again refer back to the Ophiuchus results for a discussion of halo-to-halo scatter. 

\begin{figure}[h]
    \centering
  \includegraphics[scale=0.34]{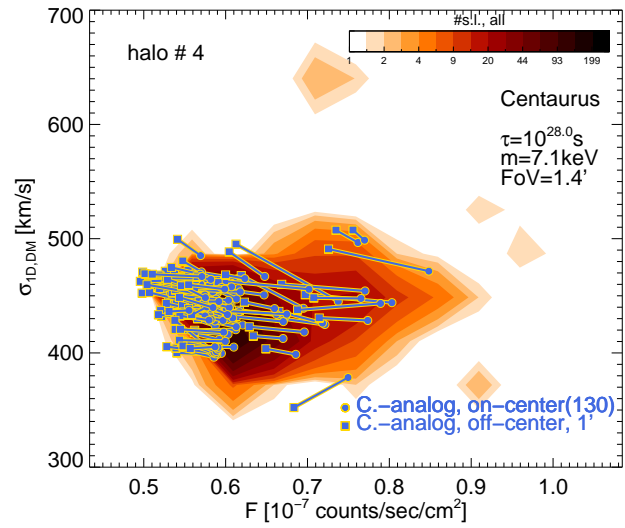}
    \caption{Repetition of Fig.~\ref{fig:XRISMcontour} for halo~4 when placed at the Centaurus redshift.}
    \label{fig:XRISMcontourCentaurus}
\end{figure}

The distribution of halo~4 sightlines prefers a flux of $0.6\times10^{-7}\mathrm{counts/sec/cm^{2}}$ with a tail maximum of $0.9\times10^{-7}\mathrm{counts/sec/cm^{2}}$. The range of allowed velocity dispersions is [380,500]~\kms, with a small minority of sightlines preferring a much higher $\sigma_\mathrm{1D,DM}\sim650$~\kms. The velocity dispersion is largely independent of flux; we anticipate from Fig.~\ref{fig:XRISMcontourPhi} that other haloes would show a weak trend, however. Finally, we note that the Centaurus $\sigma_\mathrm{1D,DM}$-selected sightlines show a clear preference for off-center fluxes to be suppressed at 10s of per~cent and $\sigma_\mathrm{1D,DM}$ to be enhanced at less than 10\%.

\end{document}